\documentclass[a4paper,11pt]{article}
\usepackage{amsmath,amsthm,amssymb}
\usepackage{graphicx}
\usepackage{mathrsfs}
\usepackage[all]{xy}
\newtheorem*{theorem*}{Theorem}

\newtheorem*{corollary*}{Corollary}

\newtheorem*{remark*}{Remark}

\newcommand{\BZ}{\mathbb{Z}}
\newcommand{\BR}{\mathbb{R}}
\newcommand{\BC}{\mathbb{C}}

\newcommand{\bD}{\mathbf{D}}
\newcommand{\bH}{\mathbf{H}}
\newcommand{\bI}{\mathbf{I}}

\newcommand{\bsigma}{\boldsymbol{\sigma}}

\newcommand{\cB}{\mathcal{B}}
\newcommand{\cC}{\mathcal{C}}
\newcommand{\cF}{\mathcal{F}}
\newcommand{\cO}{\mathcal{O}}
\newcommand{\cH}{\mathcal{H}}
\newcommand{\cL}{\mathcal{L}}

\newcommand{\ra}{\mathrm{a}}
\newcommand{\rH}{\mathrm{H}}

\newcommand{\even}{\mathrm{even}}
\newcommand{\odd}{\mathrm{odd}}
\newcommand{\NCHO}{\mathrm{NCHO}}
\newcommand{\QRM}{\mathrm{QRM}}

\renewcommand{\Im}{\operatorname{Im}}

\allowdisplaybreaks[3]

\title{Equivalence between non-commutative harmonic oscillators and two-photon quantum Rabi models}
\author{Ryosuke Nakahama\thanks{This work was supported by JST CREST Grant Number JPMJCR2113, Japan.}
\thanks{Email: ryosuke.nakahama@ntt.com} \\
\textit{NTT Institute for Fundamental Mathematics,} \\ \textit{NTT Communication Science Laboratories,} \\ \textit{Nippon Telegraph and Telephone Corporation,} \\
\textit{3-9-11 Midori-cho, Musashino-shi, Tokyo 180-8585, Japan}}
\date{\today}

\begin{document}

\maketitle

\begin{abstract}
We prove that the non-commutative harmonic oscillator on $L^2(\BR)\otimes\BC^2$ introduced by Parmeggiani and Wakayama is equivalent to the two-photon quantum Rabi model, 
and they are also equivalent to a holomorphic differential equation on the unit disk. 
The confluence process of this differential equation and the relation with the one-photon quantum Rabi model are also discussed. \medskip

\end{abstract}

\section{Introduction}

The \emph{quantum Rabi model} (QRM) is a light-matter interaction model introduced by Jaynes and Cummings \cite{JC}. Its Hamiltonian is given by the formally self-adjoint operator on $L^2(\BR)\otimes\BC^2$, 
\[ \rH_{1\QRM}^{(g,\Delta)}:=\bI\ra^\dagger\ra+g\bsigma_1(\ra+\ra^\dagger)+\Delta\bsigma_3, \]
where $g,\Delta\in\BR$, $\ra^\dagger,\ra$ are the creation and annihilation operators on $L^2(\BR)$, and $\bsigma_1,\bsigma_3\in M(2,\BC)$ are the Pauli matrices. 
Despite its simple form, QRM represents an important theoretical building block of quantized matter-field interactions and quantum information processing. 
The integrability of this model is proved by Braak \cite{B1, B2}, and this Hamiltonian has discrete spectra for all $g,\Delta$. 
Its spectral zeta functions and heat kernels are also studied by \cite{RW2, RW1, S}. 
Then as a next step, the \emph{two-photon quantum Rabi model} (or a special case of the \emph{spin-boson model}) 
\[ \rH_{2\QRM}^{(g,\Delta)}:=\bI\ra^\dagger\ra+g\bsigma_1(\ra^2+(\ra^\dagger)^2)+\Delta\bsigma_3, \]
is considered (see e.g. \cite{B3, DXBC, L, TH, XC} and references therein). This Hamiltonian has discrete spectra for $|g|<\frac{1}{2}$, 
but may have both discrete and continuous spectra for $|g|=\frac{1}{2}$. Such a phenomenon does not occur in the one-photon case, 
and these two models behave differently, despite their similar appearance.  

Independently, as a purely mathematical model, motivated by number-theoretic interests and lower bound estimates of systems of (pseudo)differential operators, 
Parmeggiani and Wakayama \cite{PW0, PW1, PW2} introduced the \emph{non-commutative harmonic oscillator} (NCHO), 
with the Hamiltonian operator on $L^2(\BR)\allowbreak\otimes\BC^2$, 
\[ \rH_\NCHO^{(\alpha,\beta)}:=\begin{bmatrix}\alpha&0\\0&\beta\end{bmatrix}\biggl(-\frac{1}{2}\frac{d^2}{dx^2}+\frac{1}{2}x^2\biggr)
+\begin{bmatrix}0&-1\\1&0\end{bmatrix}\biggl(x\frac{d}{dx}+\frac{1}{2}\biggr), \]
where $\alpha,\beta\in\BR$. If $\alpha\beta>1$, then this has discrete spectra, and if $\alpha\beta=1$, then this has continuous spectra \cite{P2014, PV}. 
The spectra of NCHO are studied mathematically also via the spectral zeta functions, e.g., the automorphic properties related to special values of the zeta 
(see \cite{IW2, KW4, O3} and references therein), and via pseudodifferential operators (see e.g., \cite{M, MP}).  
By \cite{O1, O2, RW, W}, the restriction of NCHO on the space of even or odd functions is equivalent to a Heun differential equation. 
On the other hand, by \cite{B2, KRW}, the (one-photon) QRM is equivalent to a confluent Heun differential equation, 
and NCHO is considered as a covering model of QRM via the confluence process of the Heun equation \cite{RW, W}. 
However, this confluence process seems complicated and difficult to explain reasonably. 
Indeed, in \cite{RW, W}, the explicit Hilbert space structure of the Heun picture of NCHO is not considered, 
and the correspondence of the parameters in this confluence process is not given in an accessible form. 

In this article, we prove that the eigenvalue problem of NCHO is equivalent to that of the two-photon QRM, 
and that their restriction to the space of even or odd functions is also equivalent to a holomorphic differential equation on the unit disk. 
This bridges a physical model and a mathematical model, and we expect that this gives a new number-theoretic insight into the study of QRMs (see \cite{HS, RW2, RW3, S}). 
We also discuss the confluence process of this holomorphic differential equation. 
Then we find that considering the two-photon QRM is more reasonable as a covering model of the one-photon QRM than considering NCHO. 
We note that the equivalence of NCHO and a holomorphic differential equation on the unit disk holds in a more general setting, from the viewpoint of the theory of spherical harmonics. See \cite{N}.

\section{Preliminaries}

First, let $\ra,\ra^\dagger,\rH$ be the unbounded operators on $L^2(\BR)$ given by 
\[ \ra:=\frac{1}{\sqrt{2}}\biggl(x+\frac{d}{dx}\biggr), \quad \ra^\dagger:=\frac{1}{\sqrt{2}}\biggl(x-\frac{d}{dx}\biggr), \quad 
\rH:=\ra^\dagger\ra+\frac{1}{2}=\frac{1}{2}\biggl(x^2-\frac{d^2}{dx^2}\biggr). \]
Then for $\theta\in\BR$, the unitary operator $e^{i\theta\rH}$ on $L^2(\BR)$ is given by 
\begin{align*}
e^{i\theta\rH}f(x)&=\frac{1}{\sqrt{2\pi\cos\theta}}\int_\BR e^{i(\tan\theta)(x^2+\xi^2)/2+i(\sec\theta)x\xi}\widehat{f}(\xi)\,d\xi && (\cos\theta\ne 0) \\
&=\frac{\sqrt{i}}{\sqrt{2\pi\sin\theta}}\int_\BR e^{-i(\cot\theta)(x^2+\xi^2)/2+i(\csc\theta)x\xi}f(\xi)\,d\xi && (\sin\theta\ne 0), 
\end{align*}
where $\widehat{f}(\xi):=\frac{1}{\sqrt{2\pi}}\int_\BR f(x)e^{-ix\xi}\,dx$ is the Fourier transform, if we choose the branches of $\sqrt{\cos\theta},\sqrt{\sin\theta}$ suitably. 
This has the periodicity $4\pi$ (see e.g. \cite[Corollary 4.55]{F}). 

Next, let $\cF(\BC)\subset\cO(\BC)$ be the Fock space, 
\[ \cF(\BC):=\biggl\{ f\in\cO(\BC)\biggm| \frac{1}{\pi}\int_\BC |f(w)|^2e^{-|w|^2}\,dw<\infty \biggr\}. \]
Then $L^2(\BR)$ and $\cF(\BC)$ are unitarily isomorphic by the Bargmann transform 
\begin{gather*}
\cB\colon L^2(\BR)\longrightarrow \cF(\BC), \qquad 
(\cB f)(w):=\pi^{-1/4}\int_{\BR}f(x) e^{-(x^2+w^2)/2+\sqrt{2}xw}\,dx. 
\end{gather*}
Then this satisfies 
\[ (\cB\ra f)(w)=\frac{\partial}{\partial w}(\cB f)(w), \quad (\cB\ra^\dagger f)(w)=w(\cB f)(w), \quad (\cB e^{i\theta\rH} f)(w)=e^{i\theta/2}(\cB f)(e^{i\theta}w). \]
The orthonormal basis $\Bigl\{\frac{\pi^{-1/4}}{\sqrt{m!}}(a^\dagger)^m e^{-x^2/2}\Bigr\}\subset L^2(\BR)$ is mapped to $\Bigl\{\frac{w^m}{\sqrt{m!}}\Bigr\}\allowbreak\subset \cF(\BC)$ by $\cB$ 
(see e.g. \cite[Sections 1.6, 1.7]{F}). 

Next, let $\bH,\bD\subset\BC$ be the upper half plane and the unit disk, 
\[ \bH:=\{y\in\BC\mid \Im y>0\}, \qquad \bD:=\{z\in\BC\mid |z|<1\}. \]
For $\nu>0$, let $\langle\cdot,\cdot\rangle_\nu$ be the weighted Bergman inner product on the space $\cO(\bD)$ of holomorphic functions on $\bD$ given by, 
for $F(z)=\sum_{m=0}^\infty a_mz^m$, $G(z)=\sum_{m=0}^\infty b_mz^m$, 
\begin{align*}
\langle F,G\rangle_\nu:\hspace{-3pt}&=\sum_{m=0}^\infty \frac{m!}{(\nu)_m}a_m\overline{b_m} && (\nu>0) \\
&=\frac{\nu-1}{\pi}\int_\bD F(z)\overline{G(z)}(1-|z|^2)^{\nu-2}\,dz && (\nu>1),
\end{align*}
where $(\nu)_m:=\nu(\nu+1)(\nu+2)\cdots(\nu+m-1)$, and let 
\[ \cH_\nu(\bD):=\{F(z)\in\cO(\bD)\mid \langle F,F\rangle_\nu<\infty\}. \]
Next, let $\cL_\even=\cL_{1/2}$, $\cL_\odd=\cL_{3/2}$ be the Laplace transforms 
\begin{align*}
\cL_\even&=\cL_{1/2}\colon L^2(\BR)_\even\longrightarrow \cO(\bH), & (\cL_\even f)(y)&:=\pi^{-1/4}\int_\BR f(x)e^{x^2yi/2}\,dx, \\
\cL_\odd&=\cL_{3/2}\colon L^2(\BR)_\odd\longrightarrow \cO(\bH), & (\cL_\odd f)(y)&:=\sqrt{2}\pi^{-1/4}\int_\BR f(x)xe^{x^2yi/2}\,dx, 
\end{align*}
let $\Phi_\even=\Phi_{1/2}$, $\Phi_\odd=\Phi_{3/2}$ be the integral transforms 
\begin{align*}
\Phi_\even&=\Phi_{1/2}\colon \cF(\BC)_\even\longrightarrow \cO(\bD), & (\Phi_\even f)(z)&:=\frac{1}{\pi}\int_\BC f(w)e^{z\overline{w}^2/2}e^{-|w|^2}\,dw, \\
\Phi_\odd&=\Phi_{3/2}\colon \cF(\BC)_\odd\longrightarrow \cO(\bD), & (\Phi_\odd f)(z)&:=\frac{1}{\pi}\int_\BC f(w)\overline{w}e^{z\overline{w}^2/2}e^{-|w|^2}\,dw, 
\end{align*}
and for $\nu\in\BC$, let $\cC_\nu$ be the weighted Cayley transform, 
\[ \cC_\nu\colon \cO(\bH)\longrightarrow\cO(\bD), \qquad (\cC_\nu F)(z):=(1+z)^{-\nu}F\biggl(i\frac{1-z}{1+z}\biggr), \]
so that the following diagrams commute. 
\[ \xymatrix{ L^2(\BR)_\even \ar[r]^{\cB} \ar[d]_{\cL_\even=\cL_{1/2}} & \cF(\BC)_\even \ar[d]^{\Phi_\even=\Phi_{1/2}} \\ \cO(\bH) \ar[r]_{\cC_{1/2}} & \cO(\bD), } \qquad  
\xymatrix{ L^2(\BR)_\odd \ar[r]^{\cB} \ar[d]_{\cL_\odd=\cL_{3/2}} & \cF(\BC)_\odd \ar[d]^{\Phi_\odd=\Phi_{3/2}} \\ \cO(\bH) \ar[r]_{\cC_{3/2}} & \cO(\bD). } \]
Then the compositions 
\begin{align*}
\cC_{1/2}\circ\cL_{1/2}&=\Phi_{1/2}\circ\cB\colon L^2(\BR)_\even\longrightarrow\cH_{1/2}(\bD), \\ 
\cC_{3/2}\circ\cL_{3/2}&=\Phi_{3/2}\circ\cB\colon L^2(\BR)_\odd\longrightarrow\cH_{3/2}(\bD)
\end{align*}
are the unitary isomorphisms, so that
\[ L^2(\BR)=L^2(\BR)_\even\oplus L^2(\BR)_\odd\simeq \cH_{1/2}(\bD)\oplus\cH_{3/2}(\bD). \]
These isomorphisms satisfy, for $\nu=1/2, 3/2$, 
\begin{gather*}
\cC_\nu\circ\cL_\nu\circ\biggl(\ra^\dagger\ra+\frac{1}{2}\biggr)=\biggl(2z\frac{d}{dz}+\nu\biggr)\circ\cC_\nu\circ\cL_\nu, \\
\cC_\nu\circ\cL_\nu\circ\frac{1}{2}\ra^2=\frac{d}{dz}\circ\cC_\nu\circ\cL_\nu, \qquad 
\cC_\nu\circ\cL_\nu\circ\frac{1}{2}(\ra^\dagger)^2=\biggl(z^2\frac{d}{dz}+\nu z\biggr)\circ\cC_\nu\circ\cL_\nu, 
\end{gather*}
and especially, for $f\in L^2(\BR)_\even$ $(\nu=1/2)$ or $f\in L^2(\BR)_\odd$ $(\nu=3/2)$, 
\[ (\cC_\nu\cL_\nu e^{i\theta\rH}f)(z)=e^{i\theta\nu}(\cC_\nu\cL_\nu f)(e^{2i\theta}z). \]
The orthonormal basis $\Bigl\{\frac{\pi^{-1/4}}{\sqrt{(2m)!}}(a^\dagger)^{2m}e^{-x^2/2}\Bigr\}\subset L^2(\BR)_\even$ is mapped to 
$\Bigl\{\sqrt{\frac{(1/2)_m}{m!}}z^m\Bigr\}\allowbreak\subset \cH_{1/2}(\bD)$ by $\cC_{1/2}\circ\cL_{1/2}$, 
and $\Bigl\{\frac{\pi^{-1/4}}{\sqrt{(2m+1)!}}(a^\dagger)^{2m+1}e^{-x^2/2}\Bigr\}\subset L^2(\BR)_\odd$ is mapped to \linebreak
$\Bigl\{\sqrt{\frac{(3/2)_m}{m!}}z^m\Bigr\}\subset \cH_{3/2}(\bD)$ by $\cC_{3/2}\circ\cL_{3/2}$ (see e.g. \cite[Section 4.6]{F}). 

In summary, the correspondence of each operator is given as follows. 
\[ \begin{array}{c||c|c|c|c} L^2(\BR) & \rH=\ra^\dagger\ra+\frac{1}{2} & \frac{1}{2}\ra^2 & \frac{1}{2}(\ra^\dagger)^2 & e^{i\theta\rH}f(x) \\
\cF(\BC) & w\frac{d}{dw}+\frac{1}{2} & \frac{1}{2}\frac{d^2}{dw^2} & \frac{1}{2}w^2 & e^{i\theta/2}f(e^{i\theta}w) \\
\cH_\nu(\bD) \; \bigl(\nu=\frac{1}{2},\frac{3}{2}\bigr) & 2z\frac{d}{dz}+\nu & \frac{d}{dz} & z^2\frac{d}{dz}+\nu z & e^{i\theta\nu}F(e^{2i\theta}z) \end{array} \]

\section{Equivalence between NCHO and two-photon QRM}

For $\alpha,\beta\in\BR$, $\eta\sqrt{\alpha\beta-1}\in\BR$, in \cite{RW} the $\eta$-shifted non-commutative harmonic oscillator $\rH_\NCHO^{(\alpha,\beta,\eta)}$ is defined 
as the Hamiltonian operator on $L^2(\BR)\otimes\BC^2$, 
\[ \rH_\NCHO^{(\alpha,\beta,\eta)}:=\begin{bmatrix}\alpha&0\\0&\beta\end{bmatrix}\biggl(-\frac{1}{2}\frac{d^2}{dx^2}+\frac{1}{2}x^2\biggr)
+\begin{bmatrix}0&-1\\1&0\end{bmatrix}\biggl(x\frac{d}{dx}+\frac{1}{2}\biggr)+2\eta\sqrt{\alpha\beta-1}i\begin{bmatrix}0&-1\\1&0\end{bmatrix}. \]
$\eta=0$ case is the original non-commutative harmonic oscillator introduced in \cite{PW0, PW1, PW2}. 
Next, for $g,\Delta,\varepsilon\in\BR$, let $\widetilde{\rH}_{2\QRM}^{(g,\Delta,\varepsilon)}$ be the operator on $L^2(\BR)\otimes\BC^2$ given by 
\[ \widetilde{\rH}_{2\QRM}^{(g,\Delta,\varepsilon)}:=\bI\biggl(\ra^\dagger\ra+\frac{1}{2}\biggr)+g\bsigma_1(\ra^2+(\ra^\dagger)^2)+\Delta\bsigma_3+\varepsilon\bsigma_1, \]
where $\bI=\bigl[\begin{smallmatrix}1&0\\0&1\end{smallmatrix}\bigr]$, and $\bsigma_1=\bigl[\begin{smallmatrix}0&1\\1&0\end{smallmatrix}\bigr]$, 
$\bsigma_2=\bigl[\begin{smallmatrix}0&-i\\i&0\end{smallmatrix}\bigr]$, $\bsigma_3=\bigl[\begin{smallmatrix}1&0\\0&-1\end{smallmatrix}\bigr]$ are the Pauli matrices. 
Then $\rH_{2\QRM}^{(g,\Delta,\varepsilon)}:=\widetilde{\rH}_{2\QRM}^{(g,\Delta,\varepsilon)}-\frac{1}{2}\bI$ is a special case of the spin-boson model (e.g. \cite{TH}), 
and especially if $\varepsilon=0$, then this is the two-photon quantum Rabi model (e.g. \cite{DXBC, L}). 
In addition, for $\nu>0$, let $\widetilde{\rH}_{(\nu)}^{(g,\Delta,\varepsilon)}$ be the formally self-adjoint operator on $\cH_\nu(\bD)\otimes\BC^2$ given by 
\[ \widetilde{\rH}_{(\nu)}^{(g,\Delta,\varepsilon)}:=\bI\biggl(2z\frac{d}{dz}+\nu\biggr)+2g\bsigma_1\biggl((1+z^2)\frac{d}{dz}+\nu z\biggr)+\Delta\bsigma_3+\varepsilon\bsigma_1. \]
Then these are related as follows. 

\begin{theorem*}
Let $\alpha,\beta\in\BR_{>0}$, $\eta\sqrt{\alpha\beta-1},\lambda\in\BR$, and let 
\begin{gather*}
\frac{1}{2\sqrt{\alpha\beta}}=:g, \quad -\frac{2\eta\sqrt{\alpha\beta-1}}{\sqrt{\alpha\beta}}=:\varepsilon, \\
\frac{\lambda}{2}(\alpha^{-1}+\beta^{-1})=:\mu, \quad -\frac{\lambda}{2}(\alpha^{-1}-\beta^{-1})=:\Delta, \quad 
\begin{bmatrix} \sqrt{\alpha}e^{-\pi i/4}&0\\0&\sqrt{\beta}e^{\pi i/4}\end{bmatrix}=:l. 
\end{gather*}
\begin{enumerate}
\item Let $f(x)\in L^2(\BR)\otimes\BC^2$. Then the following (a) and (b) are equivalent. 
\begin{enumerate}
\item $\rH_\NCHO^{(\alpha,\beta,\eta)}f=\lambda f$. 
\item $\widetilde{\rH}_{2\QRM}^{(g,\Delta,\varepsilon)}e^{\frac{\pi i}{4}\rH}lf=\mu e^{\frac{\pi i}{4}\rH}lf$. 
\end{enumerate}
\item Suppose $f(x)\in L^2(\BR)_\even\otimes\BC^2$, and let 
\[ F(z):=e^{\frac{\pi i}{8}}(\cC_{1/2}\cL_\even lf)(iz)=(\cC_{1/2}\cL_\even e^{\frac{\pi i}{4}\rH} lf)(z)\in\cH_{1/2}(\bD). \]
Then (a), (b) are equivalent to
\begin{enumerate}
\setcounter{enumii}{2}
\item $\widetilde{\rH}_{(1/2)}^{(g,\Delta,\varepsilon)}F=\mu F$. 
\end{enumerate}
\item Suppose $f(x)\in L^2(\BR)_\odd\otimes\BC^2$, and let 
\[ F(z):=e^{\frac{3\pi i}{8}}(\cC_{3/2}\cL_\odd lf)(iz)=(\cC_{3/2}\cL_\odd e^{\frac{\pi i}{4}\rH} lf)(z)\in\cH_{3/2}(\bD). \]
Then (a), (b) are equivalent to
\begin{enumerate}
\setcounter{enumii}{3}
\item $\widetilde{\rH}_{(3/2)}^{(g,\Delta,\varepsilon)}F=\mu F$. 
\end{enumerate}
\end{enumerate}
\end{theorem*}

Indeed, the formula (a) is rewritten as 
\[ \biggl[ \begin{bmatrix}\alpha&0\\0&\beta\end{bmatrix}\biggl(\ra^\dagger\ra+\frac{1}{2}\biggr)+\frac{1}{2}\begin{bmatrix}0&-1\\1&0\end{bmatrix}(\ra^2-(\ra^\dagger)^2)
+2\eta \sqrt{\alpha\beta-1}i\begin{bmatrix}0&-1\\1&0\end{bmatrix} \biggr]f=\lambda f. \]
Then by the definition of $l$ and $g,\varepsilon,\mu,\Delta$, we have $l^\dagger l=\bigl[\begin{smallmatrix}\alpha&0\\0&\beta\end{smallmatrix}\bigr]$ and 
\begin{gather*}
\frac{1}{2}l^{\dagger-1}\begin{bmatrix}0&-1\\1&0\end{bmatrix} l^{-1}=\frac{i}{2\sqrt{\alpha\beta}}\begin{bmatrix} 0&1\\1&0 \end{bmatrix}=gi\bsigma_1, \\
2\eta \sqrt{\alpha\beta-1}il^{\dagger-1}\begin{bmatrix}0&-1\\1&0\end{bmatrix} l^{-1}=-\frac{2\eta\sqrt{\alpha\beta-1}}{\sqrt{\alpha\beta}}\begin{bmatrix} 0&1\\1&0 \end{bmatrix}
=\varepsilon\bsigma_1, \\
\lambda l^{\dagger-1}l^{-1}=\begin{bmatrix}\lambda\alpha^{-1}&0\\0&\lambda\beta^{-1}\end{bmatrix}=\begin{bmatrix}\mu-\Delta&0\\0&\mu+\Delta\end{bmatrix}=\mu\bI-\Delta\bsigma_3. 
\end{gather*}
Hence the above equation is equivalent to 
\begin{equation}
\biggl[ \bI\biggl(\ra^\dagger\ra+\frac{1}{2}\biggr)+gi\bsigma_1(\ra^2-(\ra^\dagger)^2)+\Delta\bsigma_3+\varepsilon\bsigma_1 \biggr]lf=\mu lf. \tag{$*$}
\end{equation}
Then by 
\[ \ra e^{i\theta\rH}=e^{i\theta}e^{i\theta\rH}\ra, \qquad \ra^\dagger e^{i\theta\rH}=e^{-i\theta}e^{i\theta\rH}\ra^\dagger, \]
this is equivalent to 
\[ \biggl[ \bI\biggl(\ra^\dagger\ra+\frac{1}{2}\biggr)+g\bsigma_1(\ra^2+(\ra^\dagger)^2)+\Delta\bsigma_3+\varepsilon\bsigma_1 \biggr]e^{\frac{\pi i}{4}\rH}lf
=\mu e^{\frac{\pi i}{4}\rH}lf, \]
and we get (1). Next, when $f\in L^2(\BR)_\even$ $(\nu=1/2)$ or $f\in L^2(\BR)_\odd$ $(\nu=3/2)$, the formula ($*$) is transformed by $\cC_\nu\circ\cL_\nu$ to 
\[ \biggl[ \bI\biggl(2z\frac{d}{dz}+\nu\biggr)+2gi\bsigma_1\biggl(\frac{d}{dz}-z^2\frac{d}{dz}-\nu z\biggr)+\Delta\bsigma_3+\varepsilon\bsigma_1 \biggr](\cC_\nu\cL_\nu lf)(z)
=\mu l(\cC_\nu\cL_\nu f)(z), \]
and this is equivalent to 
\[ \biggl[ \bI\biggl(2z\frac{d}{dz}+\nu\biggr)+2g\bsigma_1\biggl((1+z^2)\frac{d}{dz}+\nu z\biggr)+\Delta\bsigma_3+\varepsilon\bsigma_1 \biggr](\cC_\nu\cL_\nu lf)(iz)
=\mu l(\cC_\nu\cL_\nu f)(iz). \]
Since we have $(\cC_\nu\cL_\nu e^{\frac{\pi i}{4}\rH}lf)(z)=e^{\frac{\pi i}{4}\nu}(\cC_\nu\cL_\nu lf)(e^{\frac{\pi i}{2}}z)
=e^{\frac{\pi i}{4}\nu}(\cC_\nu\cL_\nu lf)(iz)$, we get (2), (3). 

Theorem (1) shows that for each fixed eigenvalue $\lambda$ of $\rH_\NCHO^{(\alpha,\beta,\eta)}$, 
there exist parameters $(g,\Delta,\varepsilon)$ and an eigenvalue $\mu$ of $\widetilde{\rH}_{2\QRM}^{(g,\Delta,\varepsilon)}$ such that 
their eigenfunctions satisfy the equivalent differential equations. 
Note that this does not imply the equivalence of the Hamiltonian operators $\rH_\NCHO^{(\alpha,\beta,\eta)}$ and $\widetilde{\rH}_{2\QRM}^{(g,\Delta,\varepsilon)}$, 
since the parameter $\Delta$ of 2QRM depends on the eigenvalue $\lambda$ of NCHO. 

\begin{remark*}
\begin{enumerate}
\item A spectrum $\mu$ of $\widetilde{\rH}_{2\QRM}^{(g,\Delta,\varepsilon)}$ concerns that of $\rH_\NCHO^{(\alpha,\beta,\eta)}$ only when $|\mu|>|\Delta|$, 
since the parameters $\alpha,\beta>0$, $\lambda\in\BR$ satisfying $\frac{\lambda}{2}(\alpha^{-1}+\beta^{-1})=\mu$ and $-\frac{\lambda}{2}(\alpha^{-1}-\beta^{-1})=:\Delta$ exist only when $|\mu|>|\Delta|$. 
Especially, the non-existence of discrete spectra for $\rH_\NCHO^{(\alpha,\beta,0)}$ with $\alpha\beta=1$ \cite{P2014, PV} 
and the existence of discrete spectra for $\widetilde{\rH}_{2\QRM}^{(g,\Delta,0)}$ with $g=\frac{1}{2}$ \cite{B3, DXBC, L} do not contradict. 
In more detail, by \cite{B3, DXBC, L}, the continuous spectrum of 2QRM is positive, while the discrete spectrum is negative. 
The above theorem suggests that such a discrete spectrum $\mu$ of 2QRM for $g=\frac{1}{2}$ satisfies $|\mu|\le|\Delta|$, and does not correspond to a spectrum of NCHO for $\alpha\beta=1$, 
which has only continuous spectra by \cite{P2014, PV}. 
\item By \cite{RW}, the $\lambda$-eigenspace of $\rH_\NCHO^{(\alpha,\beta,\eta)}$ in $L^2(\BR)_\even\otimes\BC^2$ $(\nu=1/2)$ or $L^2(\BR)_\odd\otimes\BC^2$ $(\nu=3/2)$ is 2-dimensional only if 
\[ 2\eta\in\BZ, \qquad \frac{\lambda}{4}\frac{\alpha+\beta}{\sqrt{\alpha\beta(\alpha\beta-1)}}\pm\eta-\frac{\nu}{2}\in\BZ_{\ge 0}. \]
Similarly, by \cite{XC}, the $\mu$-eigenspace of $\widetilde{\rH}_{2\QRM}^{(g,\Delta,\varepsilon)}$ in $L^2(\BR)_\even\otimes\BC^2$ or $L^2(\BR)_\odd\otimes\BC^2$ is 2-dimensional only if 
\begin{equation}
\frac{\varepsilon}{\sqrt{1-4g^2}}\in\BZ,  \qquad \frac{\mu\mp\varepsilon}{2\sqrt{1-4g^2}}-\frac{\nu}{2}\in\BZ_{\ge 0}. \tag{$**$}
\end{equation}
These parameters coincide under the above theorem. 
\end{enumerate}
\end{remark*}

\section{Covering of one-photon QRM}

In this section, we consider the relation of $\widetilde{\rH}_{(\nu)}^{(g,\Delta,\varepsilon)}$ on $\cH_\nu(\bD)\otimes\BC^2$ $(\nu>0)$ 
and the one-photon asymmetric quantum Rabi model \cite{B1, B2, XZBL}
\[ \rH_{1\QRM}^{(g,\Delta,\varepsilon)}:=\bI\ra^\dagger\ra+g\bsigma_1(\ra+\ra^\dagger)+\Delta\bsigma_3+\varepsilon\bsigma_1 \]
on $L^2(\BR)\otimes\BC^2$, according to \cite[Remark 4.2]{N}. 
First, the equation $\widetilde{\rH}_{(\nu)}^{(g,\Delta,\varepsilon)}F=\mu F$, 
\[ \biggl[\bI\biggl(2z\frac{d}{dz}+\nu\biggr)+2g\bsigma_1\biggl((1+z^2)\frac{d}{dz}+\nu z\biggr)+\Delta\bsigma_3+\varepsilon\bsigma_1\biggr]F=\mu F, \]
that is, 
\[ 2\begin{bmatrix} z & g(1+z^2) \\ g(1+z^2) & z \end{bmatrix}\frac{dF}{dz}=\begin{bmatrix} \mu-\nu-\Delta & -\varepsilon-2g\nu z \\ -\varepsilon-2g\nu z & \mu-\nu+\Delta \end{bmatrix}F \]
is a Fuchsian differential equation with five regular singularities at $\Bigl\{\frac{\pm 1\pm\sqrt{1-4g^2}}{2g},\infty\Bigr\}$ for $g\ne 0,\pm\frac{1}{2}$. We put 
\[ g':=\sqrt{\nu}g, \qquad \Delta':=\frac{\Delta}{2}, \qquad \varepsilon':=\frac{\varepsilon}{2}, \qquad \mu':=\frac{\mu-\nu}{2}, \qquad w:=\sqrt{\nu}z. \]
Then the above becomes an equation 
\[ \biggl[\bI w\frac{d}{dw}+g'\bsigma_1\biggl(\biggl(1+\frac{w^2}{\nu}\biggr)\frac{d}{dw}+w\biggr)+\Delta'\bsigma_3+\varepsilon'\bsigma_1\biggr]F=\mu' F \]
on 
\[ \biggl\{ F(w)\in\cO(\sqrt{\nu}\bD) \biggm| \frac{\nu-1}{\nu\pi}\int_{\sqrt{\nu}\bD}|F(w)|^2\biggl(1-\frac{|w|^2}{\nu}\biggr)^{\nu-2}\,dw<\infty\biggr\}\otimes\BC^2. \]
Now we take the limit $\nu\to\infty$. Then this becomes an equation 
\[ \biggl[\bI w\frac{d}{dw}+g'\bsigma_1\biggl(\frac{d}{dw}+w\biggr)+\Delta'\bsigma_3+\varepsilon'\bsigma_1\biggr]F=\mu' F \]
on 
\[ \cF(\BC)\otimes\BC^2=\biggl\{ F(w)\in\cO(\BC) \biggm| \frac{1}{\pi}\int_{\BC}|F(w)|^2 e^{-|w|^2}\,dw<\infty\biggr\}\otimes\BC^2, \]
or equivalently, 
\[ \begin{bmatrix} w & g' \\ g' & w \end{bmatrix}\frac{dF}{dw}=\begin{bmatrix} \mu'-\Delta' & -\varepsilon'-g'w \\ -\varepsilon'-g'w & \mu'+\Delta' \end{bmatrix}F. \]
This has two regular singularities at $\pm g'$ and one irregular singularity at $\infty$, and hence this limit gives a confluence process of the Fuchsian equation. 
Also, this equation is equivalent to $\rH_{1\QRM}^{(g',\Delta',\varepsilon')}f=\mu' f$ via the Bargmann transform $\cB\colon L^2(\BR)\to \cF(\BC)$, 
and hence $\widetilde{\rH}_{(\nu)}^{(g,\Delta,\varepsilon)}$ is regarded as a covering model of the one-photon quantum Rabi model. 

\begin{remark*}
By \cite{KRW}, the $\mu'$-eigenspace of $\rH_{1\QRM}^{(g',\Delta',\varepsilon')}$ is 2-dimensional only if 
\[ 2\varepsilon'\in\BZ, \qquad \mu'+g^{\prime 2}\mp\varepsilon'\in\BZ_{\ge 0}. \]
In the formula $(**)$, by applying the above parameter changes, we get 
\begin{align*}
\frac{\varepsilon}{\sqrt{1-4g^2}}&=\frac{2\varepsilon'}{\sqrt{1-4g^{\prime 2}\nu^{-1}}}=2\varepsilon'(1+O(\nu^{-1})),  \\
\frac{\mu\mp\varepsilon}{2\sqrt{1-4g^2}}-\frac{\nu}{2}&=\frac{2\mu'+\nu\mp2\varepsilon'}{2\sqrt{1-4g^{\prime 2}\nu^{-1}}}-\frac{\nu}{2}
=\biggl(\mu'+\frac{\nu}{2}\mp\varepsilon'\biggr)(1+2g^{\prime 2}\nu^{-1}+O(\nu^{-2}))-\frac{\nu}{2} \\
&=\mu'+g^{\prime 2}\mp\varepsilon'+O(\nu^{-1}). 
\end{align*}
Hence by taking the limit $\nu\to\infty$, we get the degeneracy condition on $\rH_{1\QRM}^{(g',\Delta',\varepsilon')}$. 
\end{remark*}

\noindent \textbf{Acknowledgements} The author appreciates Professor M. Wakayama for a lot of helpful comments on this topic.

\end{document}